\documentclass[aps,pra,superscriptaddress,twocolumn,showpacs]{revtex4}
\usepackage[colorlinks, linkcolor=blue, citecolor=blue, urlcolor=blue, breaklinks=true]{hyperref}
\usepackage{amsmath}
\usepackage{graphicx}
\usepackage{amssymb}
\usepackage{color}
\usepackage{relsize}
\usepackage[title]{appendix}
\usepackage{amsthm}
\usepackage{subfig}

\begin{document}

\title{Singularity of relativistic vortex beam and proper relativistic observables}

\author{Yeong Deok Han}
\address{Department of Game Contents, Woosuk University, Wanju, Cheonbuk, 565-701, Republic of Korea}

\author{Taeseung Choi}
\address{Institute of General Education, Seoul Women's University, Seoul 139-774 }
\address{School of Computational Sciences, Korea Institute for Advanced Study, Seoul 130-012, Korea }
\email{tschoi@swu.ac.kr}

\author{Sam Young Cho}
\address{Centre for Modern Physics and Department of Physics, Chongqing University, Chongqing 400044, China}


\begin{abstract}

We have studied the phase singularity of the relativistic vortex beams for the two sets of relativistic operators. One includes the new spin and orbital angular momentum (OAM) operators, which is derived from the parity-extended Poincar\'e group, and the other is composed of the (usual) Dirac spin and OAM operators. The first set predicts the same singular circulation  as the nonrelativistic vortex beams. On the other hand, the second set anticipates that the singularity of the circulation is spin orientation-dependent and can be disappeared especially for relativistic paraxial electron beam with spin parallel to the propagating direction. 
These contradistinctive predictions suggest the relativistic electron beam experiment with spin-polarized electrons for the first time to answer the long-standing fundamental question, i.e., what are the proper relativistic observables, raised from the beginning of relativistic quantum mechanics since the discovery of the Dirac equation.      
     
\end{abstract}


\maketitle

\section{Introduction}

Nonrelativistic electron vortex beams carrying orbital angular momentum (OAM) have recently been studied and well-understood using paraxial approximation of the Schr\"odinger equation \cite{BliokhPR, Lloyd, Bliokh07, Uchida, Verbeeck10, Han}. The wavefunction of nonrelativistic electron vortex includes a phase singularity factor, $e^{il\phi}$, where $\phi$ is the azimuthal angle around the axis of the vortex, and it can carry orbital angular momentum of $l\hbar$ in which $l$ is an integer known as the topological charge \cite{BliokhPR}. As the energy of electron vortex beams reaches the relativistic regime with energy $200\sim 300$ keV \cite{BliokhPR, Lloyd, McMorran, Mafaheri}, the validity of the interpretation in the experiment of vortex with high energy electrons as relativistic electron vortex is questioned \cite{Rother, Bliokh11, Hayrapetyan, Birula, Barnett, BarnettReply}. 

To understand such relativistic electrons, one should use the Dirac equation \cite{Dirac}, which successfully describes relativistic electrons, instead of the Schr\"odinger equation. However, in the usual Dirac theory, spin angular momentum and orbital angular momentum of an electron are not separately conserved as in the Schr\"odinger theory. Bialynicki-Birula {\textit et al.} proved the assertion that any acceptable solutions of the Dirac equation cannot be the eigenstates of OAM using the (usual) Dirac OAM, and showed that the vortex lines continuously smeared out into all over the space for their exponential solutions, which become the standard vortex wavefunction in the nonrelativistic limit \cite{Birula}. This raised the fundamental question, whether a relativistic vortex can be generated from high energy electron beams, because of the absence of the well-defined orbital angular momentum (OAM).  
In contrast, Barnett \cite{Barnett} showed that the relativistic electron vortices with well-defined OAM and phase singularity do really exist using the so-called Foldy-Woutheysen (FW) representation \cite{FW} and the vortex charge is related to the eigenvalues of OAM as the case for nonrelativistic vortices. Barnett used separately conserved spin-like and OAM-like observables that are the FW mean spin and FW mean OAM in the original representation \cite{Barnett}. 

In fact, the controversal results of Bialynicki-Birula {\textit et al.} and Barnett in \cite{Birula, Barnett} were originated from the use of different spin and OAM operators as  relativistic operators, which was indicated by Bliokh {\it et al.} \cite{Bliokh17}. Explicitly, Bialynicki-Birula {\textit et al.} used the usual Dirac spin and OAM operators but Barnett used the FW spin and FW OAM operators. Essentially such other choices are due to a lack of understanding of relativistic operators. As an unsolved fundamental issue, to obtain a proper relativistic spin operator for massive spin $1/2$ particles has been a long-standing problem from the beginning of relativistic quantum mechanics \cite{Pryce, NW, FW, Frenkel, Chakrabarti, Gursey, Bogolubov, Ryder, Choi13, Bauke}. Many discussions have been made to suggest possible proper description of spin for massive elementary particles
till now. Since each suggested spin operator has its own strengths and weaknesses, Bauke et al \cite{Bauke} have suggested various electromagnetic environments to distinguish between the proposed relativistic spin operators experimentally. 
However, most importantly, almost all suggested spin operators would not answer on the fundamental questions, i.e., why the Dirac equation, derived from the energy-momentum relation, can naturally contain spin for massive particles, and how the suggested spin operators can explain that the Dirac equation should contain spin. 

Theoretically, the problem of determining a proper spin operator may be solved most naturally by using the space-time symmetry. Wigner showed that an elementary spin $1/2$ (Dirac) particle is a unitary representation of the $3+1$ dimensional Poincar\'e group \cite{Wigner} and his $SU(2)$ little group represents spin index. However, an explicit form of the spin operator itself was not determined. 

Recently, two of us, i.e., Choi and Cho \cite{Ours} rigorously derived the spin operator for the Dirac field that transforms covariantly under the Lorentz transformation. We call this spin operator the new spin operator to distinguish it from the other spin operators such as the Dirac and the FW mean spin operators. The new spin operator was shown to be the generators of the $SU(2)$ little group of the Poincar\'e group and admit the representation of the Poincar\'e group extended by the parity (space inversion) in which the Dirac spinor resides. Also, the Lorentz boost represented by the new spin operator can provide the representation of the parity operator from which the covariant Dirac equation is derived. 

In the view of theoretical consistency and completeness, the new spin operator representing the parity-extended Poincar\'e group is compelling as a proper relativistic spin operator. Furthermore, based on the consistent relativistic description of massive particles, the new spin operator can be effectively defined as a particle spin and an antiparticle spin operator according to the action of that spin on the particle state and the antiparticle state, respectively (Table \ref{tab:table1}). Straightforwardly, the particle spin operator was also shown to be equal to the FW mean spin operator \cite{Ours}. 

The conflicting results of two recent works in \cite{Birula, Barnett} mainly originating from using the different relativistic operators motivate us to ask whether the problem of the proper spin and its corresponding OAM operators is related to the singular behavior of relativistic vortex beams and can be experimentally determined by using relativistic vortex beams. In this study, we consider the two sets of operators, i.e., one consists of the new spin and OAM operators and the other consists of the Dirac spin and OAM operators, because only the two spin operators are derived from the partity-extended
Poincar\'e group, i.e., the space-time symmetry \cite{Ours}. While the projected operators recently studied in Ref. \cite{Bliokh07} will not be considered because the projected operators will give the same results as the Dirac operators due to that the expectation values of the projected operators are the same as the those of the corresponding Dirac operators. In order to understand relativistic electron vortex structures for the two sets of relativistic operators, our analysis will be based on the local {\it Laguerre-Gauss} \cite{Barnett} spinor wave packets of Dirac electrons in the FW representation, which is a suitable expression of experimental relativistic electron beam.





\begin{table*}[ht]
  \begin{center}
    \caption{Properties and relations of the new spin ${\bf S}_{\scriptsize{N}}$, the particle spin ${\bf S}_{\scriptsize{P}}$, and ${\bf S}_{\scriptsize{AP}}$. $L({\bf p})$ is a pure boost to give momentum ${\bf p}$. Detailed explanations are given in Sec. \ref{sec:NSP} and other notations are adapted from the explanations.}
	    \label{tab:table1}
    \begin{tabular}{l|c|c}
				\hline
		 {Properties}~~~ & \multicolumn{2}{c}{$\begin{array}{lcl}  {\bf S}_{\scriptsize{N}}&=& L({\bf p})\frac{\boldsymbol{\Sigma}}{2} L^{-1}({\bf p})   \\
	 {} &=&\frac{E}{ m} {\bf S}_{\scriptsize{D}} - \frac{{\bf p} ({\bf S}_{\scriptsize{D}} \cdot {\bf p})}{m(E+m)} + i\gamma^5 \frac{1}{m}({\bf S}_{\scriptsize{D}}\times{\bf p}). \end{array}$} \\ 
			
		{}	 & \multicolumn{2}{c}{~~~${\bf S}_{\scriptsize{N}}$ gives the $2^{nd}$ Casimir of the Poincar\'e group, neither ${\bf S}_{\scriptsize{P}}$ nor ${\bf S}_{\scriptsize{AP}}$.~~~}\\
		{}	 & \multicolumn{2}{c}{~~~All three spins satisfy the $su(2)$ algebra, i.e., $[S^i, S^j]=i\epsilon_{ijk}S^k$.}\\
		
		   \hline
       \textbf{} & \textbf{Particle} & \textbf{Antiparticle}\\ \hline
      {Hamiltonian}~~~  & $H_{\scriptsize{D}}=\boldsymbol{\alpha}\cdot{\bf p}+\beta m$  & $\tilde{H}_{\scriptsize{D}}=-\boldsymbol{\alpha}\cdot{\bf p}+\beta m $ \\   \hline
			{Spin}~~~ & $\begin{array}{lcl} {\bf S}_{\scriptsize{P}}&=& U^\dagger_{\scriptsize{FW}}\frac{\boldsymbol{\Sigma}}{2}  U_{\scriptsize{FW}}= {\bf S}_{\scriptsize{FW}} \\ 
		{}	&=&{\bf S}_{\scriptsize{D}}+ \frac{{\bf p} ({\bf p}\cdot {\bf S}_{\scriptsize{D}}) - {\bf p}\cdot{\bf p} {\bf S}_{\scriptsize{D}}}{E(E+m)} + i \beta \frac{({\bf p}\times {\boldsymbol{\alpha}})}{E} \end{array} $ &  $\begin{array}{lcl} {\bf S}_{\scriptsize{AP}}&=& U_{\scriptsize{FW}}\frac{\boldsymbol{\Sigma}}{2}  U^\dagger_{\scriptsize{FW}} \\ 
		{}&=&  {\bf S}_{\scriptsize{D}}+ \frac{{\bf p} ({\bf p}\cdot {\bf S}_{\scriptsize{D}}) - {\bf p}\cdot{\bf p} {\bf S}_{\scriptsize{D}}}{E(E+m)} - i \beta \frac{({\bf p}\times {\boldsymbol{\alpha}})}{E} \end{array}$ \\
		{}	& ${\bf S}_{\scriptsize{N}}\psi_{\scriptsize{P}}(p^\mu)= {\bf S}_{\scriptsize{P}}\psi_{\scriptsize{P}}(p^\mu)$ & 
			~~~Action on the state ${\bf S}_{\scriptsize{N}}\psi_{\scriptsize{AP}}(p^\mu)= {\bf S}_{\scriptsize{AP}}\psi_{\scriptsize{AP}}(p^\mu)$ \\
      \hline
      \end{tabular}
  \end{center}
\end{table*}

\section{Dirac spin and orbital angular momentum}
\label{sec:DSP}
In this section, for a clear comparison to the new spin and the corresponding OAM in the study of the relativistic vortex, we will briefly review the original Dirac theory \cite{Dirac}. As was introduced by Dirac \cite{Dirac}, the total angular momentum is the sum of the Dirac spin 
\begin{eqnarray}
{S}^k_{\scriptsize{D}}=\frac{\Sigma^k}{2}\equiv\left( \begin{array}{cc} \frac{\sigma^k}{2} & 0\\ 0 &\frac{\sigma^k}{2} \end{array}\right)
\end{eqnarray}
and the corresponding Dirac OAM, i.e., ${\bf r}_{\scriptsize{D}} \times {\bf p}$. This total angular momentum is constant of motion under the following Dirac Hamiltonian
\begin{eqnarray}
\label{eq:DH}
H_{\scriptsize{D}}=\boldsymbol{\alpha}\cdot {\bf p} + \beta m,
\end{eqnarray}
where ${\bf r}_{\scriptsize{D}}$ is the Dirac position operator that is the canonical operator represented by $i \boldsymbol{\nabla}_{\bf p}$, $\boldsymbol{\alpha}\cdot {\bf p}=\alpha^k p^k$ for $k=\{x, y, z\}$, and $\sigma^k$ are the Pauli matrices. We use the Einstein summation convention in which the repeated indices are summed over. For the Dirac Hamiltonian, the Dirac matrices are 
\begin{eqnarray}
\alpha^k=\left( \begin{array}{cc} 0 & {\sigma^k} \\ {\sigma^k} & 0 \end{array}\right), ~~~\beta=\left( \begin{array}{cc} 1 & 0 \\ 0 & -1 \end{array} \right)
\end{eqnarray}
in the standard representation \cite{Dirac}. Here we use natural unit $\hbar=c=1$. However, as noticed, the Dirac spin and the Dirac OAM are not separately conserved with the Dirac Hamiltonian, which is the reason why Dirac introduce the spin angular momentum. That is,  
\begin{subequations}
\begin{eqnarray}
[\; H_{\scriptsize{D}}, {\bf S}_{\scriptsize{D}} \,] &=&i \boldsymbol{\alpha}\times {\bf p}, \\
\label{eq:SHC} 
 [\; H_{\scriptsize{D}}, {\bf r}_{\scriptsize{D}} \times {\bf p} \,] &=& -i \boldsymbol{\alpha}\times {\bf p}.
\label{eq:AHC}
\end{eqnarray}
\end{subequations}

The commutator of the Dirac OAM and the Dirac Hamiltonian in Eq. (\ref{eq:AHC}) is not zero because the Dirac velocity operator $i[H_{\scriptsize{D}}, {\bf r}_{\scriptsize{D}}]=\boldsymbol{\alpha}$ is not proportional to the momentum ${\bf p}$. This suggests that the existence of the Zitterbewegung \cite{Schrodinger, Thaller}, which is a fast trembling motion first observed by Schr\"odinger, is closely related to the non-conservation of the Dirac OAM. It is known that the trembling motion of the Dirac particles will not appear when there is no interference between positive and negative energy states. Hence it will be instructive to calculate the expectation values of the Dirac orbital angular momentum either for the positive or the negative energy states. 

The Dirac Hamiltonian (\ref{eq:DH}) gives the following well-known four solutions: 
\begin{eqnarray}
\label{eq:DIRACSOLS}
u^{1}(p^\mu) &=& \frac{1}{\sqrt{2m(E+m)}}\left( \begin{array}{c} E+m \\0\\ p^z\\ p^x + i p^y \end{array}\right), \\ \nonumber
u^{2}(p^\mu) &=& \frac{1}{\sqrt{2m(E+m)}}\left( \begin{array}{c} 0\\E+m \\ p^x - i p^y \\-p^z \end{array}\right), \\ \nonumber
u^{3}(p^\mu) &=& \frac{1}{\sqrt{2m(E+m)}}\left( \begin{array}{c}  - p^z\\ - p^x - i p^y\\ E+m \\0 \end{array}\right), \\ \nonumber
u^{4}(p^\mu) &=& \frac{1}{\sqrt{2m(E+m)}}\left( \begin{array}{c} - p^x + i p^y \\p^z \\0\\E+m \end{array}\right), 
\label{eq:DHSOL}
\end{eqnarray} 
where $p^\mu=(E, {\bf p})$. 
$u^{1,2}(p^\mu)$ are the two positive energy spinors with the energy eigenvalue $+E=\sqrt{{{\bf p}\cdot{\bf p}}+m^2}$ and $u^{3,4}(p^\mu)$ are the two negative energy spinors with the energy eigenvalue $-E$. 
The eigenspinors satisfy the orthogonality 
\begin{eqnarray}
\label{eq:NES}
u^{\gamma\dagger}(p^\mu)u^{\delta}(p^\mu)=\frac{E}{m}\delta_{\gamma\delta},
\end{eqnarray}
where $\gamma$, $\delta=\{1,2,3,4\}$, $u^{\gamma\dagger}(p^\mu)$ is the Hermitian conjugate of $u^\gamma(p^\mu)$, and $\delta_{\gamma\delta}$ is the Kronecker-delta. 
The expectation value of the Dirac velocity operator $\boldsymbol{\alpha}$ either for the positive or the negative energy eigenspinors becomes
\begin{eqnarray}
u^{j\dagger}(p^\mu) \boldsymbol{\alpha} u^k(p^\mu) = \frac{\bf p}{m}\delta_{jk} \mbox{ and }
u^{m\dagger}(p^\mu) \boldsymbol{\alpha} u^n(p^\mu) = -\frac{\bf p}{m}\delta_{mn}
\end{eqnarray}
for arbitrary $j,k\in\{1,2\}$ and $m,n\in\{3,4\}$. 
As a result, the expectation values of the Dirac OAM either for the positive or the negative energy eigenspinors are
\begin{eqnarray}
&& u^{j \dagger}(p^\mu) [\; H_{\scriptsize{D}},{\bf r}_{\scriptsize{D}}\times {\bf p}\,]u^{k}(p^\mu) \\ \nonumber
& = &
u^{m \dagger}(p^\mu) [\; H_{\scriptsize{D}},{\bf r}_{\scriptsize{D}}\times {\bf p}\,]u^{n}(p^\mu)=0.
\end{eqnarray}

\section{New spin and corresponding orbital angular momentum}
\label{sec:NSP}

Recently we have derived the covariant spin operator of the parity-extended Poincar\'e group whose representation corresponds to a free massive elementary field with spin $s$ \cite{Ours}. The representation space of the parity-extended Poincar\'e group for free massive spin $1/2$ fields is the direct sum $(1/2,0)\oplus(0,1/2)$ space in which the usual Dirac particle and antiparticle spinors reside. In this section, we introduce the covariant spin operator as the new spin operator 
and the corresponding OAM in association with the new position operator.

The new spin operator is originally constructed by the generators of the Poincar\'e group, however to compare the differences between the new spin and the Dirac spin explicitly, it is convenient to represent the new spin operator in the direct sum $(1/2,0)\oplus(0,1/2)$ representation space by using the Dirac spin operator as follows \cite{Ours}
\begin{eqnarray}
\label{eq:NSP}
S^k_{\scriptsize{N}}=\frac{E}{ m} S^k_{\scriptsize{D}} - \frac{p^k ({\bf S}_{\scriptsize{D}} \cdot {\bf p})}{m(E+m)} + i\gamma^5 \frac{1}{m}({\bf S}_{\scriptsize{D}}\times{\bf p})^k,
\end{eqnarray}
where $\gamma^5=\left( \begin{array}{cc} 0 & I \\ I & 0 \end{array}\right) $ is the Dirac gamma matrix in the standard representation \cite{tong}, 
$({\bf S}_{\scriptsize{D}}\times{\bf P})^k$ are the $k$-component of ${\bf S}_{\scriptsize{D}}\times{\bf P}$ and $I$ is the 2-dimensional identity matrix. We use the upper case ${\bf P}$ for a momentum operator and the lower case ${\bf p}$ for the eigenvalue of a momentum operator.  

There were derived the following two fundamental dynamical equations for free massive spin $1/2$ particle and antiparticle from the property of parity operation \cite{Ours}
\begin{subequations}
\begin{eqnarray}
\label{eq:PDE}
\left(\gamma^\mu p_\mu -m\right) \psi_{\scriptsize{P}}(p^\mu)&=&0, \\
\left(\gamma^\mu p_\mu +m\right) {\psi}_{\scriptsize{AP}}(p^\mu)&=&0, \mbox{ respectively, }
\label{eq:APDE}
\end{eqnarray}
\end{subequations}
where $\psi_{\scriptsize{P}}(p^\mu)$ and ${\psi}_{\scriptsize{AP}}(p^\mu)$ are particle and antiparticle spinors, respectively. These two Eqs. (\ref{eq:PDE}) and (\ref{eq:APDE}) are the same as the two covariant Dirac equations for particle and antiparticle spinors, where the Dirac gamma matrices are 
\begin{eqnarray}
\gamma^0=\left( \begin{array}{cc} 1 &0\\ 0 &-1\end{array}\right)=\beta \mbox{ and } 
\gamma^k=\left( \begin{array}{cc} 0 &\sigma^k\\ -\sigma^k & 0\end{array}\right)
\end{eqnarray}
in the standard representation \cite{tong}. 

There are two positive energy and two negative energy solutions for each of Eqs. (\ref{eq:PDE}) and (\ref{eq:APDE}). Among those 8 solutions, the two particle eigenspinors are the same as $u^{1,2}(p^\mu)$ in Eq. (\ref{eq:DIRACSOLS}) that are the positive-energy solutions of the $H_{\scriptsize{D}}$ and the two antiparticle eigenspinors are the following two negative energy solutions of Eq. (\ref{eq:APDE}) as \cite{tong} 
\begin{eqnarray}
v^{1}(p^\mu) &=& \frac{1}{\sqrt{2m(E+m)}}\left( \begin{array}{c}   p^z\\  p^x + i p^y\\ E+m \\0 \end{array}\right), \\ \nonumber
v^{2}(p^\mu) &=& \frac{1}{\sqrt{2m(E+m)}}\left( \begin{array}{c}  p^x - i p^y \\-p^z \\0\\E+m \end{array}\right), 
\end{eqnarray}
which are not the negative energy eigenstates $u^{3,4}(p^\mu)$ of $H_{\scriptsize{D}}$. Then $u^{1,2}(p^\mu)$ and $v^{1,2}(p^\mu)$ satisfy the same orthogonality as $u^{j\dagger}(p^\mu)u^{k}(p^\mu)=E/m \delta_{jk}$ and $v^{j\dagger}(p^\mu)v^{k}(p^\mu)=E/m \delta_{jk}$. The four spinors $u^1(p^\mu)$, $u^2(p^\mu)$, $v^1(p^\mu)$, and $v^2(p^\mu)$ are Lorentz boosted ones from the rest spinors as 
\begin{eqnarray}
u^1(p^\mu)=e^{\gamma^5 \boldsymbol{\Sigma}\cdot {\boldsymbol{\zeta}}/2} \left(\begin{array}{c} 1 \\0\\0\\0\end{array}\right), ~
u^2(p^\mu)=e^{\gamma^5 \boldsymbol{\Sigma}\cdot {\boldsymbol{\zeta}}/2} \left(\begin{array}{c} 0 \\1\\0\\0\end{array}\right),&&\\ \nonumber
v^1(p^\mu)=e^{\gamma^5 \boldsymbol{\Sigma}\cdot {\boldsymbol{\zeta}}/2} \left(\begin{array}{c} 0 \\0\\1\\0\end{array}\right), ~
v^2(p^\mu)=e^{\gamma^5 \boldsymbol{\Sigma}\cdot {\boldsymbol{\zeta}}/2} \left(\begin{array}{c} 0 \\0\\0\\1\end{array}\right),&&
\end{eqnarray}
where $e^{\gamma^5 \boldsymbol{\Sigma}\cdot {\boldsymbol{\zeta}}/2}$ is the Lorentz boost with rapidity $\mbox{\boldmath $\zeta$}= 2\, \hat{ \mathbf{p}} \tanh^{-1} [ \sqrt{{\bf p}\cdot{\bf p}}/ (E+m) ]$. 
The new spin $S^k_{\scriptsize{N}}$ in Eq. (\ref{eq:NSP}) can also be expressed by using the following relations \cite{Ours}
\begin{eqnarray}
S^k_{\scriptsize{N}}=e^{\gamma^5 \boldsymbol{\Sigma}\cdot {\boldsymbol{\zeta}}/2}\frac{\Sigma^k}{2} e^{-\gamma^5 \boldsymbol{\Sigma}\cdot {\boldsymbol{\zeta}}/2}.
\end{eqnarray} 
Then it is easily seen that the four spinors $u^{1,2}(p^\mu)$ and $v^{1,2}(p^\mu)$ are also eigenstates of the new spin $S^k_{\scriptsize{N}}$ with the same eigenvalues of the rest spin $\Sigma^k/2$ for the rest spinors $u^{1,2}(k^\mu)$ and $v^{1,2}(k^\mu)$, where $k^\mu=(m,{\bf 0})$. 

However, the  $S^k_{\scriptsize{N}}$ is not a good observable because it is not Hermitian as seen in Eq. (\ref{eq:NSP}). This does not mean that the  $S^k_{\scriptsize{N}}$ is not a proper spin operator. In fact, the $S^k_{\scriptsize{N}}$ becomes the Hermitian particle spin operator $S^k_{\scriptsize{P}}$ and antiparticle spin operator $S^k_{\scriptsize{AP}}$ as they act on the particle states $u^{1,2}(p^\mu)$ and the antiparticle states $v^{1,2}(p^\mu)$, respectively \cite{Ours}, i.e.,
\begin{eqnarray} 
S^k_{\scriptsize{N}}u^{1,2}(p^\mu)&=&S^k_{\scriptsize{P}}u^{1,2}(p^\mu) \mbox{ and } \\ \nonumber
 S^k_{\scriptsize{N}}v^{1,2}(p^\mu)&=&S^k_{\scriptsize{AP}}v^{1,2}(p^\mu).
\end{eqnarray}
Accordingly, the explicit form of the particle and antiparticle spin operators are given as
\begin{eqnarray}
S^k_{\scriptsize{P}}&=& \frac{m}{E}e^{-\gamma^0\gamma^5\boldsymbol{\Sigma}\cdot {\boldsymbol{\zeta}}/2} \frac{\Sigma^k}{2} e^{\gamma^0\gamma^5 \boldsymbol{\Sigma}\cdot {\boldsymbol{\zeta}}/2} \\ \nonumber
&=& S^k_{\scriptsize{D}}+ \frac{p^k ({\bf p}\cdot {\bf S}_{\scriptsize{D}}) - {\bf p}\cdot{\bf p} S^k_{\scriptsize{D}}}{E(E+m)} + i \beta \frac{({\bf p}\times {\boldsymbol{\alpha}})^k}{E}, \\ \nonumber
S^k_{\scriptsize{AP}}&=& \frac{m}{E}e^{\gamma^0\gamma^5\boldsymbol{\Sigma}\cdot {\boldsymbol{\zeta}}/2} \frac{\Sigma^k}{2} e^{-\gamma^0\gamma^5 \boldsymbol{\Sigma}\cdot {\boldsymbol{\zeta}}/2} \\ \nonumber
&=& S^k_{\scriptsize{D}}+ \frac{p^k ({\bf p}\cdot {\bf S}_{\scriptsize{D}}) - {\bf p}\cdot{\bf p} S^k_{\scriptsize{D}}}{E(E+m)} - i \beta \frac{({\bf p}\times {\boldsymbol{\alpha}})^k}{E} 
\end{eqnarray}
in the momentum representation. 
Note that the particle and the antiparticle spin operators can be expressed by using the FW transformation matrix $U_{\scriptsize{FW}}(\mathbf{p})$ because
\begin{eqnarray}
\label{eq:FWUM}
\sqrt{\frac{m}{E}}e^{\gamma^0\gamma^5 \boldsymbol{\Sigma}\cdot {\boldsymbol{\zeta}}/2}=\frac{E+m+\beta \boldsymbol{\alpha}\cdot {\bf p}}{\sqrt{2E(E+m)}}=U_{\scriptsize{FW}}({\bf p})
\end{eqnarray} 
and thus the particle spin $S^k_{\scriptsize{P}}$ is straightforwardly shown to be the same with the FW mean spin operator discussed in Ref. \cite{FW, Gursey}.  

It has been shown that the $S^k_{\scriptsize{N}}$ gives Noether's conserved spin angular momentum \cite{Ours}. The conservation of the spin can be also confirmed by using the commutators between spins and the corresponding Hamiltonians. Since the antiparticle spinors $v^{1,2}(p^\mu)$ satisfy the different covariant equation (\ref{eq:APDE}) from Eq. (\ref{eq:PDE}) of the $u^{1,2}(p^\mu)$, the corresponding Hamiltonian for the $v^{1,2}(p^\mu)$ is also different from the original Dirac Hamiltonian, which is obtained as \cite{tong}
\begin{eqnarray}
\tilde{H}_{\scriptsize{D}}= -\boldsymbol{\alpha}\cdot {\bf p} + \beta m.
\end{eqnarray}
Then one can easily check that the particle and antiparticle spins are conserved because the spins commute with the corresponding Hamiltonians as follows:
\begin{eqnarray}
[H_D, S^k_{\scriptsize{P}}]=0 \mbox{ and } [\tilde{H}_{\scriptsize{D}}, S^k_{\scriptsize{AP}}]=0.
\end{eqnarray}
These facts that the particle and the antiparticle spins,  $S^k_P$ and $S^k _{AP}$, are the conserved quantities in each Hamiltonian imply that the corresponding conserved OAMs, which are respectively commuting with the corresponding Hamiltonians $H_D$ and $\tilde H_D$, can be determined through the total angular momentum. On the other hand, one can see that the Dirac OAM does not commute with both the $H_{\scriptsize{D}}$ and $\tilde{H}_{\scriptsize{D}}$. 

To obtain the commuting OAMs with the $H_{\scriptsize{D}}$ and $\tilde{H}_{\scriptsize{D}}$, it is needed to define the new position operator ${\bf R}_{\scriptsize{N}}$ in the momentum representation corresponding to the new spin operator ${\bf S}_{\scriptsize{N}}$ as \cite{Ourold}
\begin{eqnarray}
{\bf R}_{\scriptsize{N}}= e^{\gamma^5 \boldsymbol{\Sigma}\cdot{\boldsymbol{\zeta}}/2} {\bf r}_{\scriptsize{D}} e^{-\gamma^5 \boldsymbol{\Sigma}\cdot{\boldsymbol{\zeta}}/2}.
\end{eqnarray}  
The ${\bf R}_{\scriptsize{N}}$ satisfies the same commutation relations as those of the Dirac position operator, i.e, $[{R}^j_{\scriptsize{N}},R^k_{\scriptsize{N}}]=0$ (locality condition), $[R^j_{\scriptsize{N}}, S^k_{\scriptsize{N}}]=0$, and $[R^j_{\scriptsize{N}}, P^k]=i\delta_{jk}$ for $j$, $k\in\{x,y,z\}$. The new position operator acted on the particle and the antiparticle states becomes the following Hermitian particle and antiparticle position operators, respectively,
\begin{subequations}
\begin{eqnarray}
\label{eq:PPO}
R^k_{\scriptsize{P}}&=& U^\dagger_{\scriptsize{FW}}({\bf p})r^k_{\scriptsize{D}} U_{\scriptsize{FW}}({\bf p}) \\ \nonumber
&=& { r}^k_{\scriptsize{D}} +\frac{i\beta {\alpha}^k}{2E} -\frac{i \beta { p^k} (\boldsymbol{\alpha}\cdot{\bf p}) + \sqrt{{\bf p}\cdot{\bf p}} (\boldsymbol{\Sigma}\times {\bf p})^k}{2  E(E+m)\sqrt{{\bf p}\cdot{\bf p}}} \mbox{ and} \\
\label{eq:APPO}
R^k_{\scriptsize{AP}}&=& U_{\scriptsize{FW}}({\bf p})r^k_{\scriptsize{D}} U^\dagger_{\scriptsize{FW}}({\bf p}) \\ \nonumber
&=& { r}^k_{\scriptsize{D}} +\frac{i\beta {\alpha}^k}{2E} -\frac{i \beta { p}^k (\boldsymbol{\alpha}\cdot{\bf p}) - \sqrt{{\bf p}\cdot{\bf p}} (\boldsymbol{\Sigma}\times {\bf p})^k}{2 E(E+m)\sqrt{{\bf p}\cdot{\bf p}}} , 
\end{eqnarray}
\end{subequations}
 similar to the new particle and antiparticle spin operators $S^k_{\scriptsize{P}}$ and $S^k_{\scriptsize{AP}}$.

Subsequently, the velocity operators for the particle and the antiparticle are determined as 
\begin{subequations}
\begin{eqnarray}
\frac{d R^k_{\scriptsize{P}}}{dt}&=&-i[R^k_{\scriptsize{P}}, H_{\scriptsize{D}}]=\frac{p^k}{E}\frac{H_{\scriptsize{D}}}{E}, \\
\frac{d R^k_{\scriptsize{AP}}}{dt}&=&-i[R^k_{\scriptsize{AP}}, \tilde{H}_{\scriptsize{D}}]=\frac{p^k}{E}\frac{\tilde{H}_{\scriptsize{D}}}{E}, 
\end{eqnarray}
\end{subequations}
respectively. 
The particle and antiparticle spinors $u^{1,2}(p^\mu)$ and $v^{1,2}(p^\mu)$ are eigenstates of the above velocity operators with the  eigenvalue $p^k/E$ and $-p^k/E$, respectively, because $v^{1,2}(p^\mu)$ have the negative energy eigenvalue $-E$. 
This shows that there is no Zitterbewegung for the new position operators. Consequently, the particle and the antiparticle OAMs defined by ${\bf R}_{\scriptsize{P/AP}}\times {\bf p}$ are conserved by themselves. The OAM conservations of the free massive particles and antiparticles are verified by the commutation relations:
\begin{subequations}
\begin{eqnarray}
\label{eq:PAMCON}
[L^k_{\scriptsize{P}}, H_{\scriptsize{D}}] &=& \epsilon_{klm}[ R^l_{\scriptsize{P}} P^m ,  H_{\scriptsize{D}}]= 0, \\
\label{eq:APAMCON}
[L^k_{\scriptsize{AP}}, \tilde{H}_{\scriptsize{D}}] &=& \epsilon_{klm}[ R^l_{\scriptsize{AP}} P^m,  \tilde{H}_{\scriptsize{D}}]= 0,
\end{eqnarray}
\end{subequations}
where $\epsilon_{klm}$ is the Levi-Civita symbol with $\epsilon_{123}=1$.


\section{Existence of singular relativistic vortices}
\label{sec:VTXS}
In nonrelativistic case, free electron vortex states (with phase singularity) carry a well-defined OAM, which requires the conservation of the OAM \cite{Allen, BliokhPR, Lloyd}. It is natural to expect that the conserved OAM is essential also for the existence of the relativistic electron (Dirac particle) vortices. As was studied in the previous sections, the Zitterbewegung of the Dirac position operator makes the Dirac OAM not conserved as seen in Eq. (\ref{eq:AHC}). While the particle position operator shows no Zitterbewegung and as a result, gives the conserved particle OAM in Eq. (\ref{eq:PAMCON}). Therefore, the eigenstates of the particle OAM operator would compose the eigenstates of the particle Hamiltonian with a well-defined particle  OAM like those of the nonrelativistic case, but the eigenstates of the Dirac OAM do not. This raises the question "Whether the existence of the singular relativistic vortex in experiment could be a probe to proper spin and position operators?". We call this question 'Which-operator-question'. To answer the 'Which-operator-question', a specific solution for the relativistic beams is needed.




Let us first consider the particle spin and the particle OAM, which admit the vortex solutions with well-defined OAM. We assume the relativistic beam to be paraxial, which propagates mainly along $z$-direction, i.e., $|p^z| \gg |p^x|,~ |p^y|$ \cite{NotePa}. The vortex solutions expressed in terms of the eigenstates of the particle OAM can be most easily studied in the FW representation for electrons, because the particle position and the particle OAM operators are represented in the usual canonical form in the FW representation as
\begin{subequations}
\begin{eqnarray}
\label{eq:CPOS}
{r}^k&=&  U_{\scriptsize{FW}}(-i\boldsymbol{\nabla}) R^k_{\scriptsize{P}} U^\dagger_{\scriptsize{FW}}(-i\boldsymbol{\nabla})=i\partial_{p^k},  \\
\label{eq:FWAM}
L^z &=&U_{\scriptsize{FW}}(-i\boldsymbol{\nabla}) L^z_{\scriptsize{P}} U^\dagger_{\scriptsize{FW}}(-i\boldsymbol{\nabla})=-i\partial_\phi,
\end{eqnarray} 
\end{subequations}
with ${\bf P}=-i\boldsymbol{\nabla}$ in Eq. (\ref{eq:FWUM}), where $\partial_{p^k}=\partial/(\partial p^k)$, $\partial_\phi=\partial/(\partial\phi)$, and $\phi$ is the azimuthal angle of the cylindrical coordinate $(\rho,\phi,z)$ in the FW representation. 

The FW transformations of the state $\psi(x)$ and the Dirac Hamiltonian $H_{\scriptsize{D}}$ in the original representation are performed as 
\begin{subequations}
\begin{eqnarray}
\label{eq:FWST}
\psi_{\scriptsize{FW}}(x)&=&  \sqrt{\frac{m}{E}} U_{\scriptsize{FW}}(-i\boldsymbol{\nabla})\psi(x), \\
\label{eq:FWH}
H_{\scriptsize{FW}} &=&U_{\scriptsize{FW}}(-i\boldsymbol{\nabla}) H_{\scriptsize{D}} U^\dagger_{\scriptsize{FW}}(-i\boldsymbol{\nabla})\\ \nonumber
&=&\beta E,
\end{eqnarray} 
\end{subequations}
where $\psi(x)$ is the eigenstate of the Dirac Hamiltonian with the eigenvalue $E$ and ${x}=(t, {\bf x})$, $t$ is time. 
The FW transformation of the state in Eq. (\ref{eq:FWST}) differs from the FW transformation of the state in other studies \cite{FW, Gursey, Barnett} by the normalization factor $\sqrt{m/E}$, which reflects that $u^{\dagger j}(p^\mu)u^k(p^\mu)=(E/m)\delta_{jk}$ of eigenspinors of the Dirac Hamiltonian $H_{\scriptsize{D}}$ is frame-dependent but $u^{\dagger j}_{\scriptsize{FW}} u^k_{\scriptsize{FW}}=\delta_{jk}$ of the eigenspinors of the FW Hamiltonian $H_{\scriptsize{FW}} $ is frame-independent because the eigenspinors
\begin{eqnarray}
u^1_{\scriptsize{FW}}= \left(\begin{array}{c} 1 \\0\\0\\0\end{array}\right), ~~~
u^2_{\scriptsize{FW}}=\left(\begin{array}{c} 0 \\1\\0\\0\end{array}\right)
\end{eqnarray}
do not depend on the reference frame. 
Then the eigenspinors $u^1_{\scriptsize{FW}}$ and $u^2_{\scriptsize{FW}}$ (in the FW representation) are written as the same forms as  the eigenspinors given in the rest frame of the original representation, however the $u^1_{\scriptsize{FW}}$ and $u^2_{\scriptsize{FW}}$ are the eigenspinors in the moving frame with momentum ${\bf p}$ not in the rest frame. Note that the FW representation is equivalent to the original representation only for the specific momentum ${\bf p}$, i.e., not covariant under the Lorentz boost. 

In the FW representation we should consider the vortex solutions for the FW Hamiltonian $H_{\scriptsize{FW}}$ in Eq. (\ref{eq:FWH}). Barnett \cite{Barnett} showed that the expansion of the FW Hamiltonian for relativistic electrons gives the paraxial wave equation whose solution has a phase factor $e^{il\phi}$ with the eigenvalue $l$ of the canonical OAM in the FW representation. Therefore, the vortex solution in the cylindrical coordinate $(\rho, \phi,z)$ has the form \cite{REFNOTE1}
\begin{eqnarray}
\psi_{\scriptsize{FW}}(x)=e^{-iEt}\psi_{\scriptsize{FW}}(\rho,z)e^{il\phi}(au^1_{\scriptsize{FW}} + bu^2_{\scriptsize{FW}}),
\end{eqnarray}
where $|a|^2+|b|^2=1$. We consider the solution $\psi_{\scriptsize{FW}}(x)$ mono-energetic with the energy $E$ for simplicity. Then it is enough to analyze only the spatial dependence of the $\psi_{\scriptsize{FW}}(x)$, i.e., 
\begin{eqnarray}
\label{eq:FWVS}
\psi_{\scriptsize{FW}}({\bf x})=\psi_{\scriptsize{FW}}(\rho,z)e^{il\phi}(au^1_{\scriptsize{FW}} + bu^2_{\scriptsize{FW}}),
\end{eqnarray}
where ${\bf x}=(\rho, \phi, z)$.

One should be careful to calculate the expectation values in the FW representation in order to maintain the physical equivalence between the original and the FW representations for the superposition states with different momentum, because the FW representation is not covariant under the Lorentz boost that changes the momentum of the particle. The expectation value of the operator $\mathcal{O}$ at ${\bf x}$ in the original representation given by 
\begin{eqnarray}
\left\langle \mathcal{O} \right\rangle_{\bf x} \equiv \psi^\dagger({\bf x}) \mathcal{O}\psi({\bf x}),
\end{eqnarray} 
which is not simply the same as 
\begin{eqnarray}
\label{eq:FWEXPO}
\psi^\dagger_{\scriptsize{FW}}({\bf x}) \mathcal{O}^{\prime}\psi_{\scriptsize{FW}}({\bf x}),
\end{eqnarray}
where $\mathcal{O}^{\prime}=U_{\scriptsize{FW}}(-i\boldsymbol{\nabla})\mathcal{O} U^\dagger_{\scriptsize{FW}}(-i\boldsymbol{\nabla}) $ is the operator representative in the FW representation. The right expression in the FW representation is (Appendix A)
\begin{eqnarray}
\left\langle \mathcal{O} \right\rangle_{\bf x}=\psi^\dagger_{\scriptsize{FW}}({\bf x})\frac{E}{m} U_{\scriptsize{FW}}(i\overleftarrow{\boldsymbol{\nabla}}) U^\dagger_{\scriptsize{FW}}(-i\boldsymbol{\nabla})\mathcal{O}^{\prime}\psi_{\scriptsize{FW}}({\bf x}),
\end{eqnarray}
where $\overleftarrow{\boldsymbol{\nabla}}$ operates on $\psi^\dagger_{\scriptsize{FW}}({\bf x})$ to the left, 
which is clearly not the same as Eq. (\ref{eq:FWEXPO}). This relation provides nontrivial spin-orbit interaction effect in the terms of the Dirac spin and the Dirac OAM rather than the new spin and the new OAM. 
The normalized expectation value of $\mathcal{O}$ is divided by the probability amplitude $\psi^\dagger({\bf x})\psi({\bf x})$ that is also not the same as $\psi^\dagger_{\scriptsize{FW}}({\bf x})\psi_{\scriptsize{FW}}({\bf x})$ (Appendix A). 

The velocity operator corresponding to the canonical position operator ${\bf r}$ of Eq. (\ref{eq:CPOS}) in the FW representation becomes
\begin{eqnarray}
{\bf v}_{\scriptsize{FW}}= -i[{\bf r}, H_{\scriptsize{FW}}]=\frac{\bf P}{E}.
\end{eqnarray} 
This ${\bf v}_{\scriptsize{FW}}$ transforms to the following particle velocity operator ${\bf v}$ obtained from the particle position operator in the original representation,
\begin{eqnarray}
{\bf v}=-i[{\bf R}_{\scriptsize{P}}, H_{\scriptsize{D}}]=\frac{\bf P}{E}\frac{H_{\scriptsize{D}}}{E}.
\end{eqnarray}
Therefore the expectation value of the particle velocity operator at ${\bf x}$, we call the particle velocity at ${\bf x}$, is written as
\begin{eqnarray}
\label{eq:PVEL1}
\left\langle {\bf v}\right\rangle_{\bf x}=\mbox{Re}\left\{  \frac{\psi^\dagger({\bf x})\frac{\bf P}{E}\psi({\bf x})}{\psi^\dagger({\bf x}) \psi({\bf x})} \right\}.
\end{eqnarray}

To calculate further, we use the same expansion for the FW Hamiltonian in Ref. \cite{Barnett} with $E=\sqrt{P_0^2+m^2}$, i.e., 
\begin{eqnarray}
H_{\scriptsize{FW}}&=& \sqrt{{\bf p}\cdot{\bf p}+m^2} \\ \nonumber
&\approx& \sqrt{p_0^2+m^2} +\frac{(p^x)^2+(p^y)^2}{2\sqrt{p_0^2+m^2}}+ \frac{\sqrt{p_0(p^2-p_0)}}{\sqrt{p_0^2+m^2}},
\end{eqnarray}
where $p_0 \approx p^z$,
 then we obtain the following local Laguerre-Gauss (LG) solution with the form
\begin{eqnarray}
\label{eq:LGVS}
&&\psi_{\scriptsize{FW}}({\bf x}) = e^{ip_0 z} \tilde{\psi}_{\scriptsize{FW}}({\bf x})\\ \nonumber
&=&  e^{ip_0 z} \frac{1}{w^{|l|+1}(z)}\exp\left[-\frac{\rho^2}{w^2(z)}\right] L_n^{|l|}\left(\frac{2\rho^2}{w^2(z)} \right)\rho^{|l|}e^{il\phi} \\ \nonumber
&\times& \exp\left[-i(2n+|l|+1)\tan^{-1}\left(\frac{z}{z_0}\right)+i \frac{2 z \rho^2}{z_0 w(z)}\right] \\ \nonumber
&\times& (au^1_{\scriptsize{FW}} + bu^2_{\scriptsize{FW}}),
\end{eqnarray}
where $w(z)$ is the beam width, $z_0=(p_0 w(0)^2)/2$, and $\tilde{\psi}_{\scriptsize{FW}}({\bf x})$ satisfies the paraxial wave equation \cite{Barnett, BliokhPR}. In the  approximation the state varies gradually only along the $z$-axis and 
\begin{eqnarray}
\partial_{z}\psi_{\scriptsize{FW}}({\bf x})&=& p_0 \psi_{\scriptsize{FW}}({\bf x}) + e^{ip_0z} \partial_{z} \tilde{\psi}_{\scriptsize{FW}}({\bf x}) \\ \nonumber
& \approx& p\psi_{\scriptsize{FW}}({\bf x}),
\end{eqnarray}
hence the $z$ dependence of the solution can be considered solely by $e^{ip_0 z}$, i.e., the $w(z)$ can be replaced by $w(0)\equiv w$. 
We are interested in the singular behavior of the relativistic wave solutions for $\rho \to 0$, hence the region of $\rho<w/\sqrt{2}$ will be considered. However, the $\rho$ should be greater than $1/m$, the Compton wavelength, because one particle theory is not valid in the region less than the Compton wavelength in which the pair production is inevitable. Thus, in our study, we call the region of the vortex solution determined by 
\begin{eqnarray}
\label{eq:PVR}
\frac{1}{m} <  \rho < \frac{w}{\sqrt{2}}
\end{eqnarray}
as the physical vortex region for simplicity. In the physical vortex region, the wavefunction $\psi_{\scriptsize{FW}}({\bf x})$ can be written as
\begin{eqnarray}
\psi_{\scriptsize{FW}}({\bf x})\approx d_0(l) e^{ip_0 z} \frac{1}{w^{|l|+1}}\rho^{|l|}e^{il\phi}(au^1_{\scriptsize{FW}} + bu^2_{\scriptsize{FW}})
\end{eqnarray}
for the associated Laguerre polynomial 
\begin{eqnarray}
L_n^{|l|}\left(\frac{2\rho^2}{w^2} \right) = d_0(l)+ \cdots d_n \left(\frac{2\rho^2}{w^2} \right)^n,
\end{eqnarray}
where $d_0(l)$ is the function of $l$ and $d_n$ is constant. 

The physical velocity is defined actually in the original representation and thus we should use $\psi({\bf x})$ that is obtained from the inverse FW transformation of Eq. (\ref{eq:FWST}). In the physical vortex region the denominator of the velocity in Eq. (\ref{eq:PVEL1}) can be simplified as (Appendix A) 
\begin{eqnarray}
\label{eq:PDST}
\psi^\dagger({\bf x}) \psi({\bf x}) &\approx& \frac{E}{m}\psi^\dagger_{\scriptsize{FW}}({\bf x}) \psi_{\scriptsize{FW}}({\bf x}) \\ \nonumber
&\approx& \frac{E}{m} \left[d_0(l)\frac{\rho^{|l|}}{w^{|l|+1}}\right]^2.
\end{eqnarray}
And the numerator of the particle velocity $(v^x, v^y, v^z)$ at ${\bf x}$ in Eq. (\ref{eq:PVEL1}) can be obtained as (Appendix B) 
\begin{eqnarray}
&&\psi^\dagger({\bf x}) \left(\frac{p^x}{E}, \frac{p^y}{E}, \frac{p^z}{E}\right)\psi({\bf x})\\ \nonumber
&\approx& \left[d_0(l)\frac{\rho^{|l|}}{w^{|l|+1}}\right]^2 \left( -\frac{l y}{m\rho^2}, \frac{l x}{m\rho^2}, \frac{p_0}{m} \right). 
\end{eqnarray}
As a result, the particle velocity at ${\bf x}$ is given as
\begin{eqnarray}
\label{eq:PVEL}
\left\langle {\bf v}\right\rangle_{\bf x}=  \left( -\frac{l y}{m \rho^2}, \frac{l x}{m\rho^2}, \frac{p_0}{m} \right).
\end{eqnarray}
This particle velocity at $\mathbf{x}$ describes that electrons move in the $z$-direction with a spiral circular motion, which represents the singular vortex motion along the $z$-axis. This result shows that the relativistic vortex solution interpreted by the particle velocity and the particle OAM supports the singular vortex like the nonrelativistic vortex with the following circulation
\begin{eqnarray}
\label{eq:CIR}
\Gamma_{\scriptsize{P}} =\oint_{\scriptsize{C}} \left\langle {\bf v}\right\rangle_{\bf x}\cdot d{\bf l}=2\pi\frac{l}{m},
\end{eqnarray}
where $C$ is an arbitrary closed path around the $z$-axis.

\begin{figure}[t!]
\subfloat[]{
\includegraphics[height=1.3in]{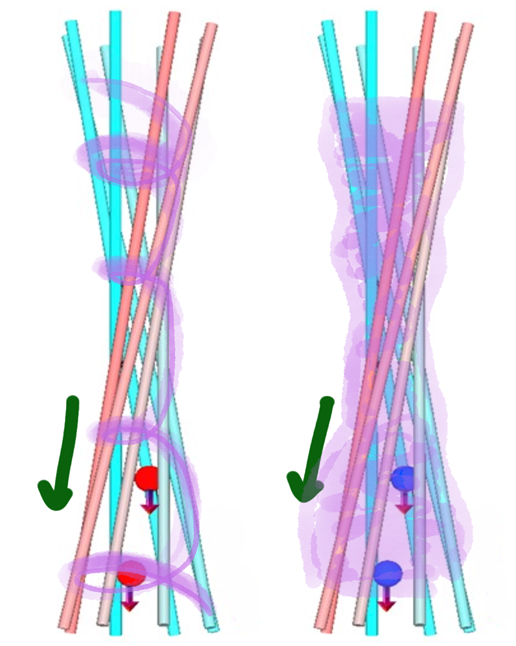}}
\subfloat[]{
\includegraphics[height=1.3in]{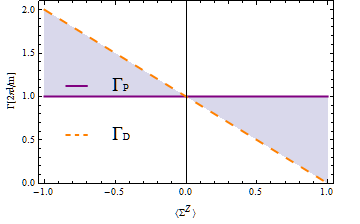}}
\caption{(Color online) (a) Schematic diagrams of spin-polarized paraxial beams for the particle velocity (left) and the Dirac velocity (right). (b) Spin-orientation dependence of the circulations of the particle velocity $\Gamma_{\scriptsize{P}}$ and the Dirac velocity $\Gamma_{\scriptsize{D}}$. This shows clearly the independence on the Dirac spin orientation for the particle velocity and the strong dependence on the Dirac spin orientation for the Dirac velocity.}
\label{fig:CAM}
\end{figure}

Next we study the singularity of the vortex solutions in Eq. (\ref{eq:LGVS}) by using the Dirac position and the Dirac OAM. The Dirac velocity at ${\bf x}$ is obtained as (Appendix B)
\begin{eqnarray}
\label{eq:DIRACVEL}
\left\langle {\bf v}_{\scriptsize{D}} \right\rangle_{\bf x} &=& \frac{\psi^\dagger({\bf x}) \boldsymbol{\alpha}\psi({\bf x})}{\psi^\dagger({\bf x})\psi({\bf x})} \\ \nonumber
&=& \frac{1}{2E \psi^\dagger_{\scriptsize{FW}}({\bf x})\psi_{\scriptsize{FW}}({\bf x})} \\ \nonumber
&\times& \left[ \left( i \boldsymbol{\nabla}\psi^\dagger_{\scriptsize{FW}}({\bf x}) \right) \psi_{\scriptsize{FW}}({\bf x})- \psi^\dagger_{\scriptsize{FW}}({\bf x}) \left( i \boldsymbol{\nabla}\psi_{\scriptsize{FW}}({\bf x}) \right) \right. \\ \nonumber
&+& \left.  \boldsymbol{\nabla}\times \left( \psi^\dagger_{\scriptsize{FW}}({\bf x}) \boldsymbol{\Sigma}\psi_{\scriptsize{FW}}({\bf x})\right)\right] \\ \nonumber
&\approx& \frac{1}{m} \left( -\frac{ly}{\rho^2}\left(1\mp \left\langle \Sigma^z \right\rangle \right), \frac{lx}{\rho^2}\left(1\mp \left\langle \Sigma^z \right\rangle \right), p_0 \right),
\end{eqnarray}
where $\mp$ corresponds to positive and negative $l$, respectively. 
Here the expectation value of the $z$-component of the Dirac spin operator $\left\langle \Sigma^z \right\rangle = \left(a^* u^{1\dagger}_{\scriptsize{FW}}+ b^* u^{2\dagger}_{\scriptsize{FW}}\right)\Sigma^z\left(a u^{1}_{\scriptsize{FW}}+ b u^{2}_{\scriptsize{FW}}\right)$ becomes approximately equal to (Appendix C)
\begin{eqnarray}
\frac{\psi^\dagger(x) \Sigma^z \psi(x)}{\psi^\dagger(x)\psi(x)}
\end{eqnarray}
for paraxial condition $|p^z|\gg |p^x|, |p^y|$, which can be also obtained from the eigenspinors in Eq. (\ref{eq:DIRACSOLS}).  Similar to the particle velocity, the Dirac velocity at $\mathbf{x}$ describes that electrons move in the $z$-direction with a spiral motion. However, in contrast to the particle velocity, the spiral motion described by the $x$- and $y$-components of the Dirac velocity depends on the expectation value of the $z$-component of the Dirac spin, i.e., the Dirac spin orientation \cite{NoteBarnett}. The circulation for the Dirac velocity 
\begin{eqnarray}
\Gamma_{\scriptsize{D}}=\oint_{\scriptsize{C}} \left\langle {\bf v}_{\scriptsize{D}}\right\rangle_{\bf x}\cdot d{\bf l}=2\pi\frac{l}{m}\left(1\mp \left\langle \Sigma^z \right\rangle \right)
\end{eqnarray} 
shows that the Dirac spin orientation determines whether a singular vortex exists or not. 
For a comparison to the circulation of the particle velocity $\Gamma_{\scriptsize{P}}$, we plot the circulation of the Dirac velocity $\Gamma_{\scriptsize{D}}$ as a function of the Dirac spin orientation $\langle \Sigma_z \rangle$ for a positive $l$ in Fig. 1(b). Figure 1(b) clearly shows that the $\Gamma_{\scriptsize{P}}$ does not depend on the Dirac spin orientation but the $\Gamma_{\scriptsize{D}}$ depends on the Dirac spin orientation. If the Dirac spin orientation is in the $xy$-plane perpendicular to the propagating direction of electron beam, $\Gamma_{\scriptsize{D}}$ is the same with $\Gamma_{\scriptsize{P}}$. However, 
compared with the $\Gamma_{\scriptsize{P}}$,
$\Gamma_{\scriptsize{D}}$ can be stronger if the angle between the Dirac spin orientation and the propagating direction of electron beam is obtuse or weaker if acute.
Especially, if the Dirac spin orientation is parallel to the propagating direction of electron beam,
the spiral circular motion of electron beam disappears, i.e., $\Gamma_{\scriptsize{D}}=0$.  Hence the spin orientation-dependent singular form of the Dirac velocity is distinct from the singularity of the particle velocity in Eq. (\ref{eq:PVEL}) (Fig. \ref{fig:CAM}).

The characteristic spin orientation-dependent property of the Dirac velocity may allow to be distinguished from the particle velocity experimentally. Especially, in order to answer 'Which-operator-question', for instance, two different setups can be considered in using spin-polarized electron beams moving in the $z$-direction in relativistic vortex experiments. One is the spin antiparallel to the propagating direction, the other is the spin parallel to the propagating direction. As we discussed in Fig. \ref{fig:CAM} (b), for the particle velocity, the two setups will give a same vortex structure independent of the spin. However, for the Dirac velocity, the two setups will give a very different observation of electron beams, i.e., the antiparallel spin gives spiral circular currents leading a vortex structure but the parallel spin gives non-spiral circular currents resulting in no vortex structure. Thus, possibly whether the non vortex structure exists or not can play a smoking-gun in distinguishing which ones can be proper relativistic operators. Consequently, distinguishable experimental observation results of relativistic vortex in such two different setups could give a clear answer for the proper relativistic observables, i.e., position, spin, and OAM. In additions, such an experimental answer on the long-standing question of the proper relativistic observables could also provide a reliable evidence to clarify whether the Zitterbewegung is a real physical effect or not. Similar results are expectable in relativistic proton vortices for the new operators using the parallel logic.

\section{Conclusion}

We have studied the singularity of relativistic electron vortex beams using two different sets of relativistic operators. The first includes the particle position, spin, and OAM operators that admit well-defined OAM $l$ for the vortex solution in the FW representation. The particle operators predict the singularity in the circulation of the relativistic LG vortex solution, which is equal to $2\pi l/m$ of the Schr\"odinger nonrelativistic vortex. The second is composed of the usual Dirac position, spin, and OAM operators by which the spin orientation-dependent singularity of the same vortex solution is anticipated. 

It was predicted that the spin seems to have little effects in the study of relativistic electron vortex beam for typical parameters in state of the art transmission electron microscopy experiments based on the estimation of the particle density $\psi^\dagger (\mathbf{x}) \psi(\mathbf{x}$ in the paraxial regime \cite{Boxem}. Actually, our study shows a similar result that the $\psi^\dagger (\mathbf{x}) \psi(\mathbf{x}$ has a considerable spin effect in Eq. (\ref{eq:LG}) (Appendix A) if the $\rho$ is smaller than the Compton wavelength, i.e., $\rho < 1/m$, while the spin effect can be negligible in $\psi^\dagger (\mathbf{x}) \psi(\mathbf{x}$ in Eq. (\ref{eq:PDST}) for the physical region, i.e., $1/m < \rho < w/\sqrt{2}$, where we have interested in. However, in sharp contrast to the behavior of the particle density $\psi^\dagger (\mathbf{x}) \psi(\mathbf{x}$ in the paraxial regime, as we discussed in Sec. \ref{sec:VTXS}, the behaviors of the particle velocity and the Dirac velocity exhibit crucial differences each other due to the spin. Then we discussed about a possible experimental setup to probe a proper set of relativistic observables based on the very different predictions from the two sets of relativistic operators for the singularity of the LG vortex solution. Especially for the paraxial electron beam with the spin parallel to the propagating direction, it could be experimentally distinguishable predictions that for the Dirac operators, no singularity- and no vortex-like motion exhibits, but for the particle operators, the singular vortex exits. Therefore, such spin-polarized relativistic electron vortex beam experiments could provide an answer on the question which relativistic observables are proper.

\section*{Acknowledgements}

 This work was supported by the Basic Science Research Program through the National Research Foundation of
Korea (NRF) funded by the Ministry of Education (2017-0342).

\section{Appendix}

\subsection{Equivalent expressions between the original and the FW representation}
 
Let us consider the Fourier transformation in the original representation
\begin{eqnarray}
\psi({\bf x})= \int d^3p e^{i{\bf p}\cdot {\bf x}}\psi({\bf p}).
\end{eqnarray}
Then the states in the FW representation are
\begin{eqnarray}
\psi_{\scriptsize{FW}}({\bf x})&=& \sqrt{\frac{m}{E}}U_{\scriptsize{FW}}(-i\boldsymbol{\nabla})\psi({\bf x})\\ \nonumber
&=& \int d^3p\sqrt{\frac{m}{E}}e^{i{\bf p}\cdot {\bf x}}U_{\scriptsize{FW}}({\bf p}) \psi({\bf p}).
\end{eqnarray}
Hence the expectation value of the operator $\mathcal{O}$ at ${\bf x}$ in the original representation becomes
\begin{eqnarray}
\left\langle \mathcal{O}\right\rangle_{\bf x}&=& \int d^3p d^3p' e^{-i{\bf p'}\cdot {\bf x}}\psi^\dagger({\bf p'}) \mathcal{O}e^{i{\bf p}\cdot {\bf x}}\psi({\bf p}) \\ \nonumber
&=&\frac{E}{m}\int d^3p d^3p' e^{-i{\bf p'}\cdot {\bf x}}e^{i{\bf p}\cdot {\bf x}} \\ \nonumber
&\times& \psi^\dagger_{\scriptsize{FW}}({\bf p'})U_{\scriptsize{FW}}({\bf p'})\mathcal{O} U^\dagger_{\scriptsize{FW}}({\bf p})\psi_{\scriptsize{FW}}({\bf p}) \\ \nonumber
&=&  \frac{E}{m} \psi^\dagger_{\scriptsize{FW}}({\bf x}) U_{\scriptsize{FW}}(i\overleftarrow{\boldsymbol{\nabla}})\mathcal{O}U^\dagger_{\scriptsize{FW}}(-i \boldsymbol{\nabla})\psi_{\scriptsize{FW}}({\bf x}) \\ \nonumber
&=& \frac{E}{m} \psi^\dagger_{\scriptsize{FW}}({\bf x}) U_{\scriptsize{FW}}(i\overleftarrow{\boldsymbol{\nabla}})U^\dagger_{\scriptsize{FW}}(-i \boldsymbol{\nabla})\mathcal{O}_{\scriptsize{FW}} \psi_{\scriptsize{FW}}({\bf x})
\end{eqnarray}
for $\mathcal{O}=U^\dagger_{\scriptsize{FW}}(-i \boldsymbol{\nabla})\mathcal{O}_{\scriptsize{FW}}U_{\scriptsize{FW}}(-i \boldsymbol{\nabla}).$

Next let us calculate the probability density at ${\bf x}$:
\begin{eqnarray}
&&\psi^\dagger({\bf x})\psi({\bf x}) \\ \nonumber
&=& \int d^3p d^3p'e^{-i{\bf p'}\cdot {\bf x}}e^{i{\bf p}\cdot {\bf x}} \\ \nonumber
&&\psi^\dagger_{\scriptsize{FW}}({\bf p'}) \frac{E}{m} U_{\scriptsize{FW}}({\bf p'}) U^\dagger_{\scriptsize{FW}}({\bf p})\psi_{\scriptsize{FW}}({\bf p}) \\ \nonumber
&=& \frac{E+m}{2m}\psi^\dagger_{\scriptsize{FW}}({\bf x}) \psi_{\scriptsize{FW}}({\bf x}) 
+ \frac{1}{2m(E+m)} \\ \nonumber
&\times&
 \left[ \right.\left(\boldsymbol{\nabla}\psi^\dagger_{\scriptsize{FW}}({\bf x}) \right)\cdot\left(\boldsymbol{\nabla}\psi_{\scriptsize{FW}}({\bf x}) \right) \\ \nonumber
&-&i \left(\boldsymbol{\nabla}\psi^\dagger_{\scriptsize{FW}}({\bf x}) \right) \cdot \boldsymbol{\Sigma}\times \left(\boldsymbol{\nabla}\psi_{\scriptsize{FW}}({\bf x}) \right) \left. \right]
\end{eqnarray}
using
\begin{eqnarray}
U_{\scriptsize{FW}}({\bf p})&=& \frac{E+m+ \beta \boldsymbol{\alpha}\cdot{\bf p}}{\sqrt{2E(E+m)}}, \\ 
U_{\scriptsize{FW}}({\bf p'}) U^\dagger_{\scriptsize{FW}}({\bf p})&=& \frac{(E+m)^2 -\beta \boldsymbol{\alpha}\cdot{\bf p'}\beta \boldsymbol{\alpha}\cdot{\bf p}}{2E(E+m)},
\end{eqnarray}
where we have used the fact that the expectation values of the odd terms, which have no diagonal elements, for $\psi_{\scriptsize{FW}}({\bf p})$ become zero. The $\psi^\dagger({\bf x})\psi({\bf x})$ for LG solution in Eq. (\ref{eq:LGVS}) approximately becomes 
\begin{eqnarray}
\label{eq:LG}
&&\psi^\dagger({\bf x})\psi({\bf x}) \\ \nonumber
&\approx& \frac{E}{m}\psi^\dagger_{\scriptsize{FW}}({\bf x})\psi_{\scriptsize{FW}}({\bf x})-i \frac{1}{2m(E+m)} \\ \nonumber
&\times& \left(\boldsymbol{\nabla}\psi^\dagger_{\scriptsize{FW}}({\bf x}) \right) \cdot \boldsymbol{\Sigma}\times \left(\boldsymbol{\nabla}\psi_{\scriptsize{FW}}({\bf x}) \right) \\ \nonumber
&\approx& \frac{E}{m} \frac{\rho^{|l|}}{w^{2(|l|+1)}}  d_0(l)^2\frac{\rho^{|l|}}{w^{2(|l|+1)}} \\ \nonumber
&+& \frac{1}{m(E+m)} p^3 l \left\langle \Sigma^\phi \right\rangle \frac{\rho^{|l|}}{w^{2(|l|+1)}}  d_0(l)^2\frac{\rho^{|l|-1}}{w^{2(|l|+1)}}  
\end{eqnarray}
for $\Sigma^\phi=-\sin{\phi}\Sigma^x+\cos{\phi}\Sigma^y$
 with the paraxial condition in the physical region. 
The second term in the last line becomes greater than the first when the $\rho$ satisfies
\begin{eqnarray}
\rho < \frac{l}{E+m} <  \frac{1}{m},
\end{eqnarray}
i.e., less than the Compton wavelength. Therefore, the final expression becomes 
\begin{eqnarray}
\label{eq:PDDFW}
\psi^\dagger({\bf x})\psi({\bf x}) &\approx&\frac{E}{m} \frac{\rho^{|l|}}{w^{2(|l|+1)}}  d_0(l)^2\frac{\rho^{|l|}}{w^{2(|l|+1)}} \\ \nonumber
& \approx& 
\frac{E}{m}\psi^\dagger_{\scriptsize{FW}}({\bf x})\psi_{\scriptsize{FW}}({\bf x}).
\end{eqnarray}

\subsection{The particle velocity and the Dirac velocity}

Let us calculate the numerator of the particle velocity operator:
\begin{eqnarray}
&&\mbox{Re}[\psi^\dagger({\bf x})\frac{\bf p}{E}\psi({\bf x})] \\ \nonumber
&=&-\frac{i}{2E}\left[ \psi^\dagger({\bf x}) \boldsymbol{\nabla} \psi({\bf x})- (\boldsymbol{\nabla}\psi^\dagger({\bf x}))
\psi({\bf x})\right]  \\ \nonumber
&=& \frac{1}{2m}\int d^3p d^3p'e^{-i{\bf p'}\cdot{\bf x}} e^{i{\bf p}\cdot{\bf x}} \left[ \right.\psi^\dagger_{\scriptsize{FW}}({\bf p'}) U_{\scriptsize{FW}}({\bf p'}){\bf p'}U^\dagger_{\scriptsize{FW}}({\bf p}) \\ \nonumber
&\times&\psi_{\scriptsize{FW}}({\bf p})   
+ \psi^\dagger_{\scriptsize{FW}}({\bf p'}) U_{\scriptsize{FW}}({\bf p'}){\bf p}U_{\scriptsize{FW}}^\dagger({\bf p}) \psi_{\scriptsize{FW}}({\bf p}) \left.\right]  \\ \nonumber
&=& \frac{1}{4mE(E+m)}\int d^3p d^3p'e^{-i{\bf p'}\cdot{\bf x}}\psi^\dagger_{\scriptsize{FW}}({\bf p'})({\bf p}+{\bf p'}) \\ \nonumber 
&\times& [(E+m)^2 +{\bf p'}\cdot{\bf p}-i{\bf p'}\cdot\boldsymbol{\Sigma}\times{\bf p}]e^{i{\bf p}\cdot{\bf x}} \psi_{\scriptsize{FW}}({\bf p})  \\ \nonumber
&\approx&  \rho^{2|l| -2}\left(\frac{d_0(l)}{w^{|l|+1}}\right)^2\left(-\frac{ly}{m}  ,  \frac{lx}{m} ,\frac{p_0}{m}\rho^{2} \right). 
\end{eqnarray}
Here we used the following calculations
\begin{eqnarray}
1. && \int d^3p d^3p'e^{-i{\bf p'}\cdot{\bf x}}\psi^\dagger_{\scriptsize{FW}}({\bf p'})[({\bf p}+{\bf p'})((E+m)^2\\ \nonumber
&+&{\bf p'}\cdot{\bf p})e^{i{\bf p}\cdot{\bf x}}\psi_{\scriptsize{FW}}({\bf p}) \\ \nonumber
&=& (E+m)^2(i   \boldsymbol{\nabla} \psi^\dagger_{\scriptsize{FW}}({\bf x}) \psi_{\scriptsize{FW}}({\bf x})-i \psi^\dagger_{\scriptsize{FW}}({\bf x})\boldsymbol{\nabla} \psi_{\scriptsize{FW}}({\bf x})) \\ \nonumber
&-&i \boldsymbol{\nabla} \psi^\dagger_{\scriptsize{FW}}({\bf x})\cdot\boldsymbol{\nabla}\boldsymbol{\nabla}\psi_{\scriptsize{FW}}({\bf x})+i\boldsymbol{\nabla}\boldsymbol{\nabla} \psi^\dagger_{\scriptsize{FW}}({\bf x}) \cdot\boldsymbol{\nabla}\psi_{\scriptsize{FW}}({\bf x}).
\end{eqnarray}
The $x$-component of the above term for the LG solution becomes 
\begin{eqnarray}
\label{eq:VELX}
&&i \partial_{x}\psi^\dagger_{\scriptsize{FW}}({\bf x})\psi_{\scriptsize{FW}}({\bf x})-i \psi^\dagger_{\scriptsize{FW}}({\bf x})\partial_{x}\psi_{\scriptsize{FW}}({\bf x})\\ \nonumber
&-&i \boldsymbol{\nabla} \psi^\dagger_{\scriptsize{FW}}({\bf x})\cdot\boldsymbol{\nabla}\partial_{x}\psi_{\scriptsize{FW}}({\bf x})+i\partial_{x}\boldsymbol{\nabla} \psi^\dagger_{\scriptsize{FW}}({\bf x}) \cdot\boldsymbol{\nabla}\psi_{\scriptsize{FW}}({\bf x}) \\ \nonumber
&=& \left(\frac{d_0(l)}{w^{|l|+1}}\right)^2\rho^{2|l|-4} \\ \nonumber
&\times& \left[ -2ly(E+m)^2 \rho^{2} - 4il^2(|l|-1) y -2l(p_0)^2 y \rho^2 \right] \\ \nonumber
&\approx& -4E(E+m)|l| y \left(\frac{d_0(l)}{w^{|l|+1}}\right)^2\rho^{2|l|-2} 
\end{eqnarray}
using the physical region condition of $\rho$. The $y$ and $z$ components are similarly calculated. And
\begin{eqnarray}
2. &-&i\int d^3p d^3p'e^{-i{\bf p'}\cdot{\bf x}}e^{i{\bf p}\cdot{\bf x}} \\ \nonumber
&\times&\psi^\dagger_{\scriptsize{FW}}({\bf p'})[({\bf p}+{\bf p'}){\bf p'}\cdot\boldsymbol{\Sigma}\times{\bf p}]\psi_{\scriptsize{FW}}({\bf p}) \\ \nonumber
&=&-\boldsymbol{\nabla}\psi^\dagger_{\scriptsize{FW}}({\bf x}) \cdot \boldsymbol{\Sigma}\times{\boldsymbol{\nabla}}\boldsymbol{\nabla}\psi_{\scriptsize{FW}}({\bf x}) \\ \nonumber
&+& \boldsymbol{\nabla}\boldsymbol{\nabla}\psi^\dagger_{\scriptsize{FW}}({\bf x}) \cdot\boldsymbol{\Sigma}\times{\boldsymbol{\nabla}}\psi_{\scriptsize{FW}}({\bf x}).
\end{eqnarray}
The $x$-component of the above term (for $l>0$) becomes
\begin{eqnarray}
&-&\boldsymbol{\nabla}\psi^\dagger_{\scriptsize{FW}}({\bf x}) \cdot \boldsymbol{\Sigma}\times{\boldsymbol{\nabla}}\partial_{x}\psi_{\scriptsize{FW}}({\bf x}) \\ \nonumber
& + & \partial_{x}\boldsymbol{\nabla}\psi^\dagger_{\scriptsize{FW}}({\bf x}) \cdot\boldsymbol{\Sigma}\times{\boldsymbol{\nabla}}\psi_{\scriptsize{FW}}({\bf x}) \\ \nonumber
&=&\left\langle \Sigma^x \right\rangle \left(\frac{d_0(l)}{w^{|l|+1}}\right)^2 \rho^{2l-2}\\ \nonumber
&\times& [-2l(l-1) p_0  (\cos^2{\phi}-\sin^2{\phi}) +2p_0 l^2 \rho^{2l-2}] \\ \nonumber
&-& \left\langle \Sigma^y \right\rangle \left(\frac{d_0(l)}{w^{|l|+1}}\right)^2\rho^{2l-2} 4l(l-1)p_0 \cos{\phi}\sin{\phi} \\ \nonumber
&+&4l^2(l-1) \left\langle \Sigma^z \right\rangle \left(\frac{d_0(l)}{w^{|l|+1}}\right)^2 \rho^{2l-3}\sin{\phi}. 
\end{eqnarray}
This term can also be ignored for the physical vortex region, when compared to the term in Eq. (\ref{eq:VELX}) with order $E(E+m)\rho^{2l-1}$ because this term is order of $p_0 \rho^{2l-2}$. The $y$ and $z$ term are similarly calculated.

Next let us calculate the numerator of the Dirac velocity:
\begin{eqnarray}
&&\psi^\dagger({\bf x})\boldsymbol{\alpha}\psi({\bf x})\\ \nonumber
&=&\int d^3p d^3p' e^{-i{\bf p'}\cdot{\bf x}} e^{i{\bf p}\cdot{\bf x}}\frac{E}{m} \\ \nonumber
&\times& \psi^\dagger_{\scriptsize{FW}} ({\bf p'}) U_{\scriptsize{FW}}({\bf p'}) {\boldsymbol{\alpha}} U^\dagger_{\scriptsize{FW}} ({\bf p})\psi_{\scriptsize{FW}}({\bf p}) \\ \nonumber
&=& \frac{1}{2m} \left[ \right. i\boldsymbol{\nabla}\psi^\dagger_{\scriptsize{FW}} ({\bf x})\psi_{\scriptsize{FW}} ({\bf x}) - \psi^\dagger_{\scriptsize{FW}} ({\bf x})i\boldsymbol{\nabla}\psi_{\scriptsize{FW}} ({\bf x}) \\ \nonumber
 &+& \boldsymbol{\nabla}\times \left(\psi^\dagger_{\scriptsize{FW}} ({\bf x})\boldsymbol{\Sigma} \psi_{\scriptsize{FW}} ({\bf x}) \right) \left.\right]
\end{eqnarray}
using only non-vanishing term $\beta(E+m) [  \boldsymbol{\alpha}\cdot{\bf p'}\boldsymbol{\alpha} + \boldsymbol{\alpha} \boldsymbol{\alpha}\cdot{\bf p}] $ from $(E+m +\beta \boldsymbol{\alpha}\cdot{\bf p'})\boldsymbol{ \alpha} (E+m -\beta \boldsymbol{\alpha}\cdot{\bf p})$.
The first term that is equal to the FW velocity in the FW representation becomes
\begin{eqnarray}
&&\frac{1}{2m} \left[ i\boldsymbol{\nabla}\psi^\dagger_{\scriptsize{FW}} ({\bf x})\psi_{\scriptsize{FW}} ({\bf x}) - \psi^\dagger_{\scriptsize{FW}} ({\bf x})i\boldsymbol{\nabla}\psi_{\scriptsize{FW}} ({\bf x}) \right] \\ \nonumber
& \approx &\left(\frac{d_0(l)}{w^{|l|+1}}\right)^2\rho^{2|l|-2} \left(-\frac{l y}{m}, \frac{lx}{m} ,  \frac{p_0}{m} \left(\frac{d_0(l)}{w^{|l|+1}}\right)^2\rho^{2}\right).
\end{eqnarray} 
And the second term$\times 2m$ becomes
\begin{eqnarray}
&&\boldsymbol{\nabla}\times \left(\psi^\dagger_{\scriptsize{FW}}({\bf x}) \boldsymbol{\Sigma} \psi_{\scriptsize{FW}}({\bf x}) \right) \\ \nonumber
&=& \left(\frac{d_0(l)}{w^{|l|+1}}\right)^2\rho^{2|l|-2} \\ \nonumber
&\times& \left(2 |l| y \left\langle \Sigma^z \right\rangle , -2 |l| x \left\langle \Sigma^z \right\rangle, 2|l|(-y\left\langle \Sigma^x \right\rangle +x \left\langle \Sigma^y \right\rangle)\right).
\end{eqnarray}
Note that $\partial_{z} \psi^\dagger_{\scriptsize{FW}}({\bf x}) \Sigma^{x,y} \psi_{\scriptsize{FW}}({\bf x}) =0$. 
Finally we obtain the Dirac velocity in Eq. (\ref{eq:DIRACVEL}).

\subsection{The spin expectation value $\left\langle \Sigma^z \right\rangle$}

Note that 
\begin{eqnarray}
\left\langle \Sigma^z \right\rangle &=&  \left(a^* u^{1\dagger}_{\scriptsize{FW}}+ b^* u^{2\dagger}_{\scriptsize{FW}}\right)\Sigma^z\left(a u^{1}_{\scriptsize{FW}}+ b u^{2}_{\scriptsize{FW}}\right) \\ \nonumber
&=& |a|^2-|b|^2.
\end{eqnarray}
The following relation is obtained
\begin{eqnarray}
\frac{\psi^\dagger({\bf x}) \Sigma^z \psi({\bf x}) }{\psi^\dagger({\bf x}) \psi({\bf x}) } \approx \frac{\psi^\dagger_{\scriptsize{FW}}({\bf x}) \Sigma^z \psi_{\scriptsize{FW}}({\bf x}) }{\psi^\dagger_{\scriptsize{FW}}({\bf x}) \psi_{\scriptsize{FW}}({\bf x}) }=|a|^2-|b|^2
\end{eqnarray}
by using 
\begin{eqnarray}
&&(E+m +\beta \boldsymbol{\alpha}\cdot {\bf p'} )\Sigma^z (E+m -\beta \boldsymbol{\alpha}\cdot {\bf p} ) \\ \nonumber
&=&(E+m)^2 \Sigma^z -i(E+m)\epsilon_{zik}\beta \alpha^k (p^i+p^{'i})\\ \nonumber 
&+&p^{'z} \boldsymbol{\Sigma}\cdot {\bf p}-{\bf p'}\cdot{\bf p} \Sigma^z 
+ p^{z} \boldsymbol{\Sigma}\cdot {\bf p'}, 
\end{eqnarray}
 and
\begin{eqnarray}
&&\psi^\dagger({\bf x}) \Sigma^z \psi({\bf x}) \\ \nonumber
 &=&\frac{E}{m} \int d^3p d^3p' e^{-i{\bf p'}\cdot{\bf x}}e^{i{\bf p}\cdot{\bf x}} \\ \nonumber
&\times& \psi^\dagger_{\scriptsize{FW}}({\bf p'}) U_{\scriptsize{FW}}({\bf p'})\Sigma^z U^\dagger_{\scriptsize{FW}}({\bf p})\psi_{\scriptsize{FW}}({\bf p}) \\ \nonumber
&=& \frac{E}{m} \psi^\dagger_{\scriptsize{FW}}({\bf x}) \Sigma^z \psi_{\scriptsize{FW}}({\bf x}) ,
\end{eqnarray}
and Eq. (\ref{eq:PDDFW}).

\subsection{The expectation value of the $z$-component of the operator ${\bf r}\times \boldsymbol{\alpha}$}

The expectation value of  the $z$-component of the operator ${\bf r}\times \boldsymbol{\alpha}$ is calculated as follows
\begin{eqnarray}
&&\langle (x \alpha^y - y \alpha^x) \rangle \\ \nonumber
&=& \int d^3x \psi^\dagger (\bf x) (x \alpha^y - y \alpha^x) \psi({\bf x}) \\ \nonumber
&=& \int d^3 p \psi^\dagger_{\scriptsize{FW}} U_{\scriptsize{FW}}({\bf p}) (x \alpha^y - y \alpha^x) U^\dagger_{\scriptsize{FW}}({\bf p})\psi_{\scriptsize{FW}}.
\end{eqnarray}
The off-diagonal term of the expectation value for the $\psi_{\scriptsize{FW}}$ in Eq. (\ref{eq:FWVS}) becomes zero, so non-zero contributing terms of 
$U_{\scriptsize{FW}}({\bf p}) (x \alpha^y - y \alpha^x) U^\dagger_{\scriptsize{FW}}({\bf p})$ are 
\begin{eqnarray} \nonumber
&&i \beta \frac{1}{2E} \left[ \right.  \alpha^y \frac{p^x}{E}\frac{2E+m}{E(E+m)} \boldsymbol{\alpha}\cdot{\bf p} + {\alpha^y \alpha^x} -i \alpha^y \boldsymbol{\alpha}\cdot {\bf p} x   \\ 
&-& \frac{2E+m}{E(E+m)}\frac{\alpha^x p^y}{E} \boldsymbol{\alpha}\cdot {\bf p}-\alpha^x \alpha^y +i \alpha^x \boldsymbol{\alpha}\cdot {\bf p}y \left. \right ] \\ \nonumber
&+& \frac{i\beta \boldsymbol{\alpha}\cdot {\bf p}}{2E(E+m)} \left[ \right.  \frac{2E+m}{E^2}\alpha^y p^x + \frac{\alpha^y p^x}{E} -i (E+m) \alpha^y x \\ \nonumber
&-& \frac{2E+m}{E^2}\alpha^x p^y -  \frac{\alpha^x p^y}{E} +i (E+m) \alpha^x y \left. \right] . \\ \nonumber
&=& \beta \left[\frac{xp^y -yp^x}{E}+ \frac{\Sigma^z}{E} +\frac{\boldsymbol{\Sigma}\cdot{\bf p}-\Sigma^z {\bf p}\cdot{\bf p}}{E^2(E+m)} \right]. 
\end{eqnarray}
using ${\bf r}=i \boldsymbol{\nabla}_{\bf p}$. Hence the expectation value of $  ({\bf r}\times \boldsymbol{\alpha})^z $ becomes $\frac{1}{E}(l+2 \langle S^z_{\scriptsize{D}}\rangle +2( \langle {\bf S}_{\scriptsize{D}}\rangle\cdot{\bf p}p^z -\langle S^z_{\scriptsize{D}}\rangle{\bf p}\cdot{\bf p}) /(p_0^2(p_0+m))  )$, which is twice of Eq. (20) of Ref. \cite{Bliokh11}.

Note that the above gives $\frac{1}{E}(l+2 \langle S^z_{\scriptsize{D}}\rangle)$ for the paraxial approximation. However, the expectation value at ${\bf x}$ gives the Dirac spin-orbit interaction effect even though the average value of $p^x$ and $p^y$ over all space are zero for the paraxial LG solution in Eq. (\ref{eq:LG}), because $\psi^\dagger_{\scriptsize{FW}}(x)p^{x,y} \psi_{\scriptsize{FW}}(x)\neq 0$. 

\newpage



\begin{thebibliography}{100}
\bibitem{BliokhPR} K. Y. Bliokh, I. P. Ivanov, G. Guzzinati, L. Clark, R. Van Boxem, A. Béché, R. Juchtmans, M. A. Alonso, P. Schattschneider, F. Nori and J. Verbeeck, Phys. Rep. {\bf 690}, 1 (2017), and references therein. 
\bibitem{Lloyd} S. M. Lloyd, M. Babiker, G. Thirunavukkarasu, and J. Yuan, Rev. Mod. Phys. {\bf 89}, 035004 (2017), and references therein.
\bibitem{Bliokh07} K. Y. Bliokh, Y. P. Bliokh, S. Savel'ev, and F. Nori, Phys. Rev. Lett. {\bf 99}, 190404 (2007).
\bibitem{Uchida} M. Uchida and A. Tonomura, Nature (London) {\bf 464}, 737 (2010).
\bibitem{Verbeeck10} J. Verbeeck, H. Tian, and P. Schattschneider, Nature (London) {\bf 467}, 301 (2010).
\bibitem{Han} Y. D. Han and T. Choi, Phys. Lett. A {\bf 381}, 1335 (2017).
\bibitem{McMorran} B. J. McMorran, A. Agarwal, I. M. Anderson, A. A. Herzing, H. J. Lezer, 
J. J. McClelland, and J. Unguris, Science {\bf 331}, 192 (2011).
\bibitem{Mafaheri} E. Mafakheri, A. H. Tavabi, P.-H. Lu, R. Balboni, F. Venturi,
C. Menozzi, G. C. Gazzadi, S. Frabboni, A. Sit, R. E. DuninBorkowski, E. Karimi, and V. Grillo, Appl. Phys. Lett. 110,
093113 (2017). 
\bibitem{Rotter} A. Rotter and K. Scheerschmidt, Ultramicroscopy, {\bf 109} 154 (2009).
\bibitem{Bliokh11} K. Y. Bliokh, M. R. Dennis, and F. Nori, Phys. Rev. Lett. {\bf 107}, 174802 (2011). 
\bibitem{Hayrapetyan} A. G. Hayrapetyan, O. Matula, A. Aiello, A. Surzhykov, and S. Fritzsche, Phys. Rev. Lett. {\bf 112}, 134801 (2014).
\bibitem{Birula} I. Bialynicki-Birula and Z. Bialynicka-Birula, Phys. Rev. Lett. {\bf 118}, 114801 (2017). 
\bibitem{Barnett} S. M. Barnett, Phys. Rev. Lett. {\bf 118}, 114802 (2017).
\bibitem{BarnettReply} I. Bialynicki-Birula and Z. Bialynicka-Birula, Phys. Rev. Lett. {\bf 119}, 029501 (2017); S. M. Barnett, {\it ibid.} {\bf 119}, 029502 (2017). 

\bibitem{Dirac} P. A. M. Dirac, Proc. R. Soc. Lond. A {\bf 117}, 610 (1928).
\bibitem{FW} L. L. Foldy and S. A. Wouthuysen, Phys. Rev. {\bf 78}, 29 (1950).
\bibitem{Bliokh17} K. Y. Bliokh, M. R. Dennis, and F. Nori, Phys. Rev. A {\bf 96}, 023622 (2017).
\bibitem{Pryce} M. H. L. Pryce, Proc. R. Soc. Lond. A {\bf 195}, 62 (1948). 
 %
 \bibitem{NW} T. D. Newton and E. P. Wigner, Rev. Mod. Phys. {\bf 21}, 400 (1949).

%

%
 \bibitem{Frenkel} J. Frenkel, Z. Physik {\bf 37}, 243 (1926).

%
 \bibitem{Chakrabarti} A. Chakrabarti, J. Math. Phys. {\bf 4}, 1215 (1963).


%

%
 \bibitem{Gursey} F. G\"ursey, Phys. Lett. {\bf 14}, 330 (1965).

%

%
 \bibitem{Bogolubov} N. N. Bogolubov, A. A. Logunov, and I. T. Todorov, \textit{Introduction to Axiomatic Quantum Field Theory}
   (W. A. Benjamin, 1975).


%
 \bibitem{Ryder} L. H. Ryder, Gen. Relat. Grav. {\bf 31}, 775 (1999).

%
 \bibitem{Choi13} T. Choi, J. Korean Phys. Soc. {\bf 62}, 1085 (2013).
\bibitem{Bauke} H. Bauke, S. Ahrens, C. H. Keitel, and R. Grobe, Phys. Rev. A {\bf 89}, 052101 (2014).
\bibitem{Wigner} E. P. Wigner, Ann. of Math. {\bf 40}, 149 (1939).
\bibitem{Ours} T. Choi and S. Y. Cho, arXiv:1807.06425.
\bibitem{Schrodinger} E. Schr\"odinger, Sitz. Preuss. Kaad. Wiss. Phys.-Math. Kl. {\bf 24}, 418 (1930).
\bibitem{Thaller} B. Thaller, \textit{The Dirac Equation} (Springer-Verlag, Berlin, 1992).
\bibitem{tong}
D. Tong \textit{Quantum Field Theory} (http:\slash \slash www.damtp.cam.ac.uk\slash user\slash tong\slash qft.html) (2006).
\bibitem{Ourold}  T. Choi and S. Y. Cho, arXiv:1410.0468.
\bibitem{Allen} L. Allen, M. W. Beijersbergen, R. J. C. Spreeuw, and J. P. Woerdman, Phys. Rev. A {\bf 45}, 8185 (1992).
\bibitem{NotePa} In our paraxial approximation, the spin-orbit interaction effects are included (see Appendix D). 
\bibitem{REFNOTE1} Bialynicki-Birula et al. \cite{Birula} showed that $[L_{\scriptsize{D}}^z, H_{\scriptsize{D}}]\psi=0$ leads no vortex solution for relativistic electrons because $[L_{\scriptsize{D}}^z, H_{\scriptsize{D}}]^2=0$ becomes $(p^x)^2+(p^y)^2=0$, which provides the Laplace equation not the paraxial equation, where $L_{\scriptsize{D}}^z$ is the $z$-component of the Dirac OAM, ${\bf x}_{\scriptsize{D}}\times {\bf p}$. However, the null commutator in Eq. (\ref{eq:PAMCON}) for the $z$-component of the particle OAM does not derive $(p^x)^2+(p^y)^2=0$, because $[X^i_{\scriptsize{P}}, \alpha^j]\neq  0$ as can be seen from Eq. (\ref{eq:PPO})
unlike $[x^i_{\scriptsize{D}}, \alpha^j]=0$ of the Dirac position operator for $i$, $j$ $\in \{x,y,z\}$.

\bibitem{NoteBarnett} The same results both for the FW and the Dirac velocity are obtained for the state in Eq. (16) of Ref. \cite{Barnett}. 
\bibitem{Boxem} R. Van Boxem, J. Verbeeck, and B. Partoens, EPL (Europhysics Letters) {\bf 102}, 40010 (2013).



\end{thebibliography}

\end{document}